\documentclass[a4paper,12pt]{article}
\usepackage{graphicx}
\usepackage{graphics}
\usepackage{amssymb}
\usepackage{amsmath}
\usepackage{a4wide}
\begin{document}

\begin{titlepage}

\begin{center}
~\\
\vspace{4cm}~\\
{\LARGE \bf The Constraint on FCNC Coupling of the Top Quark with a Gluon from
$ep$~-~Collisions}\\

\vspace{2cm} A.A.~Ashimova~\footnote{E-mail:~ashimova\_aa$@$mail.ru} \\
{\it Moscow State University, 119992 Moscow, Russia} \\[2mm]
S.R.~Slabospitsky~\footnote{E-mail:~Sergey.Slabospitsky$@$ihep.ru}\\
{\it  State Research Center\\
Institute for High Energy Physics,\\
Protvino, Moscow Region 142281, Russia}

\end{center}

\vspace{3cm}
\begin{abstract} \noindent
Using the constraint on the single top production cross-section obtained
at the HERA collider, $\sigma(ep \to e \, t \, X)$, we evaluate an upper
limit on coupling constant of the anomalous top quark interaction with a
gluon via flavor-changing neutral current: $
|\kappa_{tgq}/\Lambda|\lesssim
0.4~\mbox{TeV}^{-1}$, BR$(t \to gq) < 13 \% $
\end{abstract}
\end{titlepage}

\newpage
\section*{\bf Introduction}

In different extensions of the Standard Model (SM) the anomalous
interactions can lead to significant modification of the top quark
production mechanisms and appearance of the top rare decays~\cite{Beneke:2000hk}.
Thereby, the precision measurements of the top quark properties and
its production mechanisms provide a possibility to obtain an
information on new physics beyond the SM. Of special interest it
is the study of the top quark anomalous interactions via
Flavour-Changing Neutral Currents (FCNC).

In the SM there are no direct FCNC transitions $t\to
u(c)$. Only~~''loop'' contributions can make them possible. As a
result such processes are strongly suppressed within the SM: BR$(t \to
(\gamma,g,Z) + u(c))<{\cal{O}}(10^{-10})$ \cite{fcnc-SM}. However, various
extensions of the SM could lead to a huge enhancement of FCNC
processes~\cite{dH, SUSY, exotic}.

Search for the anomalous FCNC interactions of the top quark were performed
at the Tevatron~\cite{Abe:1997fz},
HERA~\cite{fcnc-HERA,Aktas:2003yd,Chekanov:2003yt},
 and LEP-2~\cite{lep-2} colliders.
The  present constraints
on the top quark anomalous couplings in terms of branching ratio (BR) are
presented in the Table below.

\begin{table}[h]
  \centering
  \caption{\it The present constraints on top quark anomalous couplings}
\smallskip
\label{T3}
\begin{tabular}{|c|c|c|}
  \hline
  decay  & BR & collider  \\
  \hline
  \hline
  $t\to q\gamma$ & $<0.7\%$ & HERA \cite{Aktas:2003yd, Chekanov:2003yt}\\
  $t\to qZ$      & $<7\%$   & LEP 2 \cite{lep-2}\\
  $t\to qg$      & $<8\%$  & Tevatron + {\small th.} \cite{Tevatron} \\
  \hline
\end{tabular}
\end{table}

Search for the top production at HERA collider in $ep$-collisions,
\begin{equation}
    e\,p \to e \, t \, X \, ,
\label{epetX}
\end{equation}
was
done by the H1~\cite{Aktas:2003yd}  and ZEUS~\cite{Chekanov:2003yt}
  collaborations at the
center-mass energy of $\sqrt{s}=318$~GeV. Obviously that only single
top production is possible at this energy. In this case the SM
dominant process is $ ep\to\nu t \bar{b} X$.
However, the production cross-section for this process at HERA energy is very small
($\sigma(ep \to tX)<1$~fb~\cite{SMtopprod}) and far beyond an experimental
observation.
On the other hand the anomalous FCNC process could contribute to the single
 top production
at HERA energies. Therefore, a study of such reaction at HERA provide a
good possibility to put a constraint on FCNC interactions.

\begin{figure}[p]
\centering  \includegraphics[width=5cm,clip]{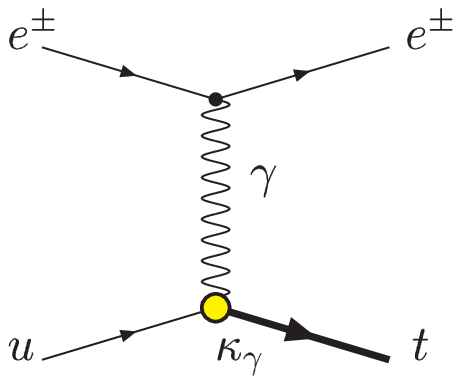}\\
\caption{\small\it Single top production via FCNC coupling with
photon (or $Z$-bozon) }\label{r2-2}
\end{figure}

The both  H1 and  ZEUS collaborations searched for  the top quark
production in reaction~(\ref{epetX}).
No evidence of the top production was found. The resulted constraint on
the single top production cross-section is given
below~\cite{Aktas:2003yd, Chekanov:2003yt}:
\begin{equation}\label{zbest}
    \sigma(e p \to e\, t \, X) < 0.225\;\;{\rm pb} \qquad ({\rm at} \;\;
95\%~~{\rm CL})
\end{equation}
The conventional interpretation of this data
assumes that the top quark anomalous FCNC interactions
could occur with a photon or $Z$-boson only (see Fig.~\ref{r2-2}).
As a result from the experimental upper limit~(\ref{zbest})
the constraint on the value of anomalous
coupling was evaluated~\cite{Aktas:2003yd, Chekanov:2003yt}. In terms of BR this constraint
 is as follows:
\begin{eqnarray}
{\rm BR}(t \to q \gamma) <0.7 \%
\label{constr1}
\end{eqnarray}

\begin{figure}[p]
\centering \includegraphics[width=12cm,height=8cm,clip]{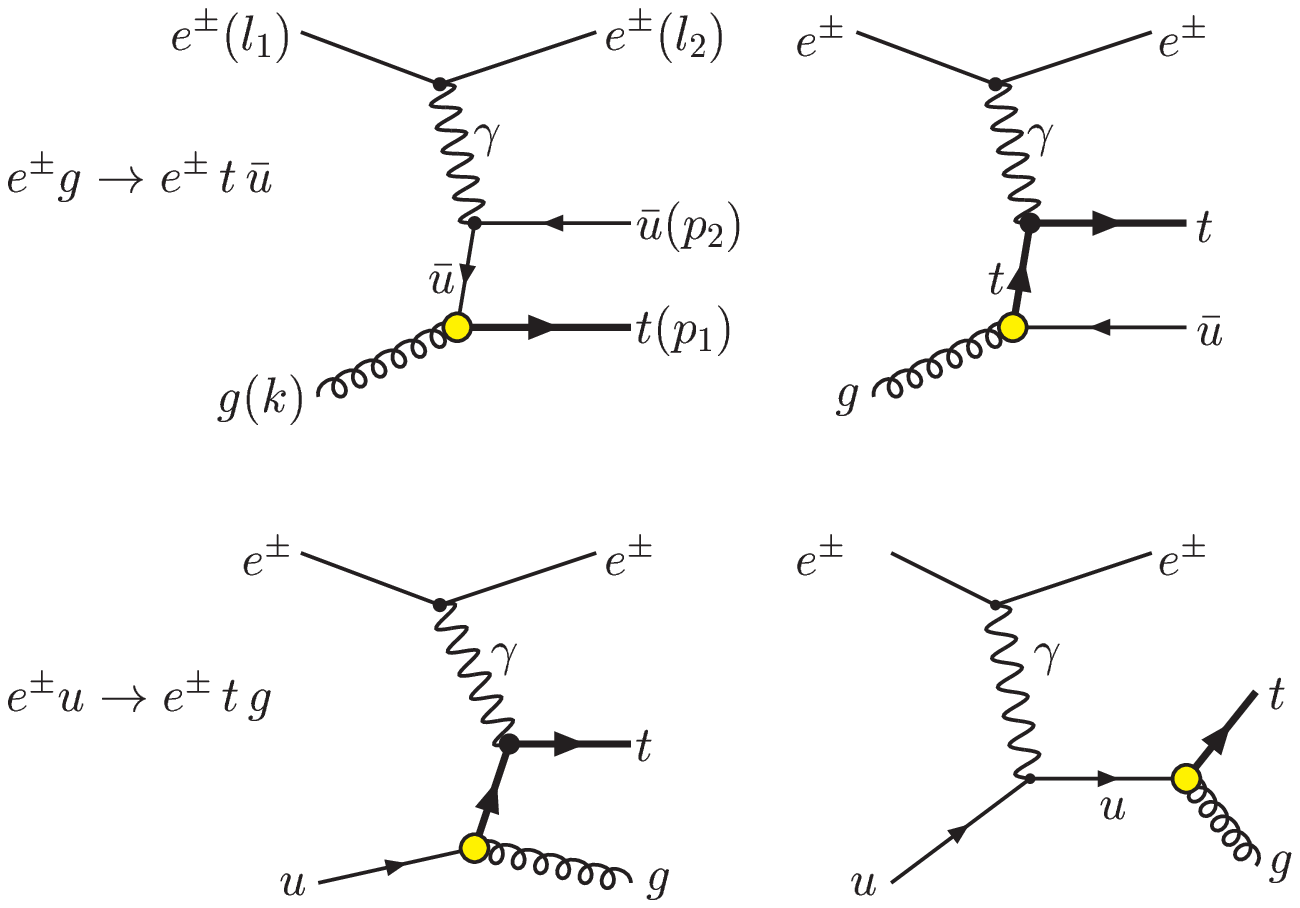}\\
\bigskip
\caption{\small\it The Feynman diagrams of the single top
production in $ep$-collisions  due to FCNC with a
gluon}\label{rD}
\end{figure}

In the present article we consider an additional FCNC process that contribute
to the single top-quark production at HERA energies. Indeed, in $ep$-collisions
the single top could be produced also due to anomalous
interaction with a gluon, i.e. $tgq$ vertex, as it is shown on fig.
\ref{rD}. Thereby using the upper limit on
the production cross-section~(\ref{zbest}) we evaluate
a constraint on anomalous FCNC coupling of the top
quark with a gluon. Therein lies the basic idea of our analysis.

\section{\bf Phenomenological Lagrangian for FCNC}

We do not know which type of new physics will be responsible for a
future deviation from the SM predictions. However, the top quark
couplings can be parameterized in a model independent way by an
effective Lagrangian~\cite{Beneke:2000hk} (see also~\cite{tait}
for the first discussion of the $t \,c \,g$-operator).

In our article we consider the top quark anomalous interactions
with gluon and photon. However for the completeness we also
present here the expression for the $t$-quark FCNC interaction
with $Z$-boson. Actually  the width of $t\to Zq$ decay is used in
the Section \ref{LSec}. Thus the effective phenomenological
Lagrangian have a form as follows:

\begin{eqnarray}\label{L}
 {\cal  L}
 = &-& 2g_{s}\sum \frac{\kappa_{g}}{\Lambda}\bar{t}\sigma_{\mu\nu}
          T^{a}(g_{L}P_{L}+g_{R}P_{R})uq^{\nu}G^{\mu}_{a} \nonumber\\
   &-& 2e\sum \frac{\kappa_{\gamma}}{\Lambda}\bar{t}\sigma_{\mu\nu}
               (\gamma_{L}P_{L}+\gamma_{R}P_{R})uq^{\nu}A^{\mu}  \\
   &-& \frac{e}{\sin 2\vartheta_{W}}\sum \kappa_{Z}\bar{t}\gamma^{\mu}(z_{L}P_{L}+z_{R}P_{R})uZ_{\mu}+h.c.,\nonumber
\end{eqnarray}
where $t^{a}$ are the Gell-Mann matrices satisfying
$Tr(t^{a}t^{b})=\delta^{ab}/2$; $\kappa_{g}$ and $\kappa_{\gamma}$
define the strength of the FCNC with a gluon and a photon
respectively (real and positive values); $\Lambda = 1$~TeV is a
new physics cutoff; $e$ is the electric charge; $\vartheta_{W}$ is
the Weinberg angle; $\sigma^{\mu\nu}=
\frac{1}{2}(\gamma^{\mu}\gamma^{\nu}-\gamma^{\nu}\gamma^{\mu})$;
$P_{R,L} = \frac{1}{2}(1\pm\gamma^{5})$;
$g^{2}_{L}+g^{2}_{R}=\gamma^{2}_{L}+\gamma^{2}_{R}=1$; $q^{\nu}$
is the momentum of corresponding gauge boson.

The top-quark decay widths into one of the light quarks ($u$ or
$c$) with a gluon, photon and $Z$-boson have the form:
\begin{equation}\label{Gg}
\Gamma(t\rightarrow
gq)=\frac{4}{3}\alpha_{s}m^{3}_{t}\left(\frac{\kappa_{g}}{\Lambda}\right)^{2}
\end{equation}

\begin{equation}\label{Gph}
\Gamma(t\rightarrow \gamma q)=\alpha
m^{3}_{t}\left(\frac{\kappa_{\gamma}}{\Lambda}\right)^{2}
\end{equation}

\begin{equation}\label{GZ}
 \Gamma(t\rightarrow Zq)=\frac{\alpha\kappa^{2}_{Z}m^{3}_{t}}{8sin^{2}2\vartheta_{W}M^{2}_{Z}}
 \left(1-\frac{M^{2}_{Z}}{m^{2}_{t}}\right)^{2}
\left(1+2\frac{M^{2}_{Z}}{m^{2}_{t}}\right)
\end{equation}\\

\section{\bf The matrix elements squared } \label{metx}
Firstly, we re-calculate the matrix element squared for $2 \to 2$ process:
$eu \to et$ (see the digram on Fig.~\ref{r2-2}). This expression is used in our
analysis.
The explicit form~\cite{belyaev} is given below:
\begin{eqnarray}
 &|M|^{2}& = 4e^{4}\left(\frac{1}{Q^{2}}\right)^{2}\left( \frac{\kappa_{\gamma}}
{\Lambda}
\right)^2\times  \label{m2-2}\\
 &~& \left\{2m^{2}_{e}\hat t(2\hat s+\hat
t)-2m^{2}_{e}m^{4}_{t}-\hat t[ 2\hat s(\hat s-m^{2}_{t})+
m^{4}_{t}]-\hat t^{2}(2\hat s-m^{2}_{t})\right\}\nonumber
\end{eqnarray}
where $\hat{s}$ and $\hat{t}$ are usual Mandelstam variables, $Q^2
= |\hat{t}|$ is a photon transferred momentum squared, $m_e$ and
$m_t$ are the masses of the electron and $t$-quark. Note, that we
keep non-zero electron mass, because this is essential for the
further analysis.

One can obtain an expression for the production cross-section of
the hard $2 \to 2$ process (corresponding to matrix element
squared~(\ref{m2-2})) by the integration on $Q^2$ from $Q^2_{min}$
to $Q^2_{max}$:
\begin{eqnarray}\label{sigma}
\hat\sigma (Q^{2}_{min})\quad &= & C \left[ B \ln
\left( \frac{Q^2_{max}}{Q^2_{min}} \right)
-A(Q^2_{max}-Q^2_{min}) \right] \nonumber\\
  &-& 2Cm^{2}_{e}\left[2\hat s\ln\left(\frac{Q^2_{max}}{Q^2_{min}}\right)
- (Q^2_{max}-Q^2_{min})\right]\\
  &+& 2Cm^{2}_{e}m^{4}_{t}\left(\frac{1}{Q^2_{min}}-\frac{1}{Q^2_{max}}\right),
\nonumber
\end{eqnarray}
where
$$C = \frac{\pi e^{4}}{2\hat s(\hat s-m^{2}_{e})}
\left(\frac{\kappa_{\gamma}}{\Lambda}\right)^2,
\qquad  A = 2\hat s-m^{2}_{t},  \qquad  B = 2\hat s(\hat s-m^{2}_{t})+m^4_{t}$$

The single top production in $ep$-collisions  due
to anomalous FCNC interaction with a gluon is described by two
subprocesses  with a gluon
or $u$($c$)-quark from the initial proton (see the diagrams on Fig.~\ref{rD}):
\begin{eqnarray}
  e  g \to e t q, \;\; q = u \;\; {\rm or} \;\; c \label{dia3_gl}\\
  e  q \to e t g,  \;\; q = u \;\; {\rm or} \;\; c \label{dia3_up}
\end{eqnarray}
Note, that the matrix elements squared for these two processes are related
to each other by crossing. Therefore, we present an exact expression
for $|M|^2$ for the first subprocess~(\ref{dia3_gl}):
\begin{equation}\label{m2-3}
    |T|^{2} = \left|T_{1}+T_{2}\right|^{2}
\end{equation}

The exact expressions are given below:
\begin{eqnarray*}
|T_{1}|^{2} &=& A_{0}A^{2}_{1}(kp_{2}) \left\{(f_{1}l_{1})(p_{1}l_{2})+(f_{1}l_{2})
(p_{1}l_{1})
      - (p_{1}f_{1})\left[(l_{1}l_{2})+\frac{q^{2}}{2}\right]\right\}\\
 &~& \quad -2A_{0}A^{2}_{1}m^{2}_{t}(kp_{2})\left[(l_{1}l_{2}) + q^{2}\right];\\
|T_{2}|^{2} &=& A_{0}A^{2}_{2}(kp_{1}) \left\{(f_{2}l_{1})(p_{2}l_{2})+(f_{2}l_{2})
(p_{2}l_{1})
          -(p_{2}f_{2})\left[(l_{1}l_{2})+\frac{q^{2}}{2}\right]\right\};
\end{eqnarray*}

\begin{eqnarray*}
  \frac{T^{+}_{1}T_{2}+T^{+}_{2}T_{1}}{A_{0}A_{1}A_{2}} &=&
4(p_{1}l_{1})(p_{1}l_{2})(kp_{2})^{2}+
4(p_{2}l_{1})(p_{2}l_{2})(kp_{1})^{2}+4(kl_{1})(kl_{2})(p_{1}p_{2})^{2} \\
&-& 2(p_{1}p_{2})\left\{2(f_{3}l_{1})(kl_{2})+2(f_{3}l_{2})(kl_{1})-
(kf_{3})(l_{1}l_{2})\right\} \\
&+& 2m^{2}_{t}(kp_{2})\left\{(p_{2}l_{1})(kl_{2})+(p_{2}l_{2})(kl_{1})
-(kp_{2})(l_{1}l_{2}\right\}\\
&-& m^{2}_{t}\left\{2(f_{3}l_{1})(kl_{2})+2(f_{3}l_{2})(kl_{1})-(kf_{3})q^{2}
\right\}\\
&+& 2m^{2}_{t}(kl_{1})(kl_{2})(p_{1}p_{2});
\end{eqnarray*}
where
\[A_{0}= 64 \left[\frac{4}{3} \frac{\kappa_{g}}{\Lambda}\alpha_{s}e^{2}\right]^{2},
 A_{1}= -\frac{1}{2(kp_{2})+m^{2}_{t}}, \quad
A_{2}=-\frac{1}{2(kp_{1})-m^{2}_{t}}\] \vspace{1pt}
\[f_{1}= 2(kp_{2})p_{2}+m^{2}_{t}k, f_{2}=2(kp_{1})p_{1}-m^{2}_{t}k, f_{3}=
(kp_{1})p_{2}+(kp_{2})p_{1}.\]
here $l_1(l_2)$ are 4-momentum of the initial (final) electron,
$p_1$ and $p_2$ are 4-momenta of the top- and $\bar u$-quarks, and $k$ is
the initial gluon momentum.
The calculation of the color coefficients is trivial and they are
included to $A_0$.

For numerical calculations the evaluated matrix elements squared are
incorporated in the event generator TopReX~\cite{toprex}.

\section{\bf Kinematics of the considered processes}\label{kinem}

The cross-section of the considered processes depends essentially
on the kinematical limit of the gauge boson transferred momentum
$Q_{min}$. This limit is determined by the additional requirements
on kinematics of the reaction and the detector parameters. For
some reason the authors of the experimental paper
\cite{Chekanov:2003yt} did not mention the limit on $Q_{min}$ in
an explicit form.

In order to calculate the theoretical value of the cross-section,
$\sigma^{\rm th}$, we must imply a correct limit on $Q_{min}$.
This value can be estimated from the following considerations. The
experimental analysis for $2 \to 2$ process was made under the
assumption that single top is produced due to anomalous coupling
with a photon. It is reasonable that the theoretical cross-section
also depends on the $Q_{min}$ kinematical limit and anomalous
coupling with a photon. Therefore by using the value of the
anomalous coupling one can compare theoretical cross-section with
experimental one and can determine the unknown value of
$Q^2_{min}$.

Thus we get the value of $Q^2_{min}$ from the following condition:
\begin{equation}\label{trebq2}
  \sigma^{\rm th}_{2\to 2}(ep \to etX)= \left(\frac{\kappa_\gamma}{\Lambda}\right)^2\times
  f^{ep\to etX}_{2\to 2}(Q^2_{min})\leqslant \sigma^{exp}=0.225\;\;\;{\rm  pb}
\end{equation}
Here $\sigma^{\rm th}_{2\to 2}$ is the total cross-section for
$ep\to etX$ process (with subprocess $eu \to et$) calculated using
the expression (\ref{sigma}), $\sigma^{exp}$ is the experimental
limit on the single top quark production from the HERA
data~\cite{Chekanov:2003yt}. The upper limit on anomalous coupling
of the top quark with a photon was taken from the same
article~\cite{Chekanov:2003yt}: $ \kappa^{H}_{tu\gamma} < 0.174$.

Note that the authors of the experimental
work~\cite{Chekanov:2003yt} used different normalization for
phenomenological Lagrangian:
$${\cal  L^{H}} =
ee_t~\bar{t}~\frac{\sigma_{\mu\nu}q^{\nu}}{\Lambda^{H}}u~\kappa^{H}_{tu\gamma},
$$ where $e$ is the electric charge, $e_t$ is the charge of up
quarks and $\Lambda^{H} = m_t = 175~{\rm GeV}$.

The relation between $\kappa^{H}_{tu\gamma}$ and $\kappa_\gamma$
used in our Lagrangian~(\ref{L}) looks as follows:
\begin{equation}\label{kappa2kappa_HERA}
\kappa_\gamma =
\frac{e_t}{\sqrt{2}}\frac{\Lambda}{\Lambda^{H}}\frac{1}{\sqrt{\gamma^2_L
+ \gamma^2_R}}\kappa^{H}_{tu\gamma}
\end{equation}
Note that we take for the new physics cutoff $\Lambda = 1~{\rm
TeV}$ and by our convention $\gamma^2_L + \gamma^2_R = 1$ (see
Eq.~(\ref{L})). As a result the upper limit on the anomalous
coupling in terms of the Lagrangian~(\ref{L}) equals:
\begin{equation}\label{kappa_gamma}
\kappa_\gamma = 0.468.
\end{equation}


The single top production cross-section in $ep$-collsions was
evaluated within the parton model:
\begin{equation}\label{sigma2-2}
    \sigma^{\rm th}(e^{\pm} \, p \, \to \, e^{\pm} \, t \, X)
 = \int\limits_{m^2_t/s}^{1} dxf(x,\mu^{2})\hat{\sigma}(x,\mu^{2}),
\end{equation}
where $x$ is the part of the proton momentum carried away by a
parton; $\hat{\sigma}(x,\mu^{2})$ is the cross-section for the
hard subprocess; $f(x,\mu^{2})$ is the parton distribution
function, and $\mu = m_t$ is the factorization scale.

Using (\ref{trebq2}) we obtain
\begin{equation}\label{qlim}
 Q^{2}_{min} \geqslant 0.08 \;\; {\rm GeV}^{2}
\end{equation}

Further in our analysis of the single top production due to
anomalous coupling with a {\bf gluon} we implied the very same
limit on $Q^2_{min}$ (\ref{qlim}).


\section{\bf The constraint on FCNC coupling of the top-quark with a gluon \label{LSec}}

In the present study we compare the experimental
constraint~(\ref{zbest}) with the total top-quark production
cross-section in $ep$-collisions (evaluated with the matrix
elements squared~(\ref{m2-3})).

\begin{equation}\label{2-3}
    ep\to etfX,
\end{equation}

Thus the single top quark production in $ep$-collisions is
possible not only via FCNC with a photon, but also due to
anomalous interactions with a gluon. This idea gives us an
opportunity to obtain the direct constraint on FCNC coupling of
the top quark with a gluon from HERA experimental data
(\ref{zbest}).

In order to find the constraint on gluon coupling $\kappa_{g}$ we
use the following relation:

\begin{equation}\label{**}
  \sigma^{th}(ep\to etfX)\times BR(t\to bW^{+}) \leqslant
 \sigma^{exp}=0.225\;\;\;{\rm  pb},
\end{equation}
where $$\sigma^{\rm th}_{2\to 3}(ep \to etfX)=
\left(\frac{\kappa_g}{\Lambda}\right)^2\times F^{ep\to etfX}_{2\to
3}(Q^2_{min}).$$

 Here we take into account the possible
modification of the branching ratio of the top decay in the SM
mode $t\to bW^{+}$. As it was mentioned above the top quark
anomalous interactions could lead to the appearance of additional
top decay channels. So the probability of the SM top decay would
change accordingly:

\begin{equation}\label{BR}
BR(t\to bW^{+}) = \frac{\Gamma(t\to
bW^{+})}{\Gamma(t\to bW^{+}) + \Gamma(t\to Zq)+
\Gamma(t\to gq)}, \vspace{10pt}
\end{equation}

To obtain an upper limit on the anomalous coupling with a gluon
we assume  the absence of the top-quark FCNC interaction with a photon,
i.e. $\kappa_{t\gamma u,c}=0$. In addition the different cases of the top
interactions with a gluon and $Z$-boson are explored:

\begin{itemize}
    \item[1.] $\kappa_{tZu,c}=0$, $\kappa_{tgc}=0$,
$\kappa_{tgu}\neq 0$,\\
the simplest case, i.e. gluon - top - $u$-quark interaction;
    \item[2.] $\kappa_{tZu,c}=0$, $\kappa_{tgc}=\kappa_{tgu}\neq 0$.\\
here the FCNC transitions with a gluon to $u$- or $c$-quark are
considered;
\end{itemize}

Two additional cases include the anomalous interaction with $Z$-boson:
\begin{itemize}
  \item[3.] $\kappa_{tZu,c}\neq 0$, $\kappa_{tgc}=0$, $\kappa_{tgu}\neq 0$.
    \item[4.] $\kappa_{tZu,c}\neq 0$, $\kappa_{tgc}=\kappa_{tgu}\neq 0$.
\end{itemize}

According to all assumptions (1.-4.) the constraints on the
top quark  FCNC couplings with a gluon, $u$-, $c$-quarks and the
corresponding BR of the anomalous top decays are
calculated.
The results of the analysis are presented in Table \ref{T4}.

\begin{table}[h]
  \centering
  \caption{\it The constraints on the top quark FCNC coupling with a
gluon from the HERA data $(\kappa_{t\gamma q}=0)$}
\smallskip
\label{T4}
\begin{tabular}{|c|c|c|c|c|}
\hline
   &  &  $\kappa_{tgu}/\Lambda$, TeV$^{-1}$ & $\kappa_{tgc}/\Lambda$,
 TeV$^{-1}$ & BR$(t \to gq)$ \\
\hline \hline
  1  & $\kappa_{tZq}=0$, $\kappa_{tgc}=0$~~~  & 0.393 & --- &  6.6\% \\
\hline
  2  & $\kappa_{tZq}=0$, $\kappa_{tgu}=\kappa_{tgc}$  & 0.396 & 0.396 & 12.6\% \\
\hline \hline
  3  & $\kappa_{tZq}\neq 0$, $\kappa_{tgc}=0$~~~  & 0.408 & ---  & 6.2\% \\
\hline
 4 & $\kappa_{tZq}\neq 0$, $\kappa_{tgu}=\kappa_{tgc}$  & 0.410 & 0.410 & 11.8\% \\
\hline
\end{tabular}
\end{table}

\parindent=1cm

The best constraint on FCNC coupling with a gluon from
$ep$-collision is obtained supposing the interaction with an
$u$-quark only. When the interaction with a charm quark is also
taken into account the constraints on couplings do not change
greatly but the BR increases almost twice. It is explained by the
small  contribution of the $c$-quarks to proton's PDF and
appearance of two decay modes of the top-quark. One also can see
that the contribution of $Z$-boson modifies the results slightly
and can be neglected.

Finally the derived constraint on the anomalous coupling
of the top FCNC interaction with a gluon and
the corresponding BR are given below:

\begin{equation} \label{result}
|\kappa_{tgq}/\Lambda|\lesssim
0.4~\mbox{TeV}^{-1}, \quad {\rm BR}(t \to gq) < 13 \%
\end{equation}

\section{\bf Conclusions}

Very recently at the Tevatron experiment \cite{Tevatron} better
constraint ($3\div10$ times less) on the $\kappa_g$, the upper
limit on the anomalous coupling of the top quark with a gluon, was
obtained.

Nevertheless we wish to indicate that the HERA results for the
single top production provides the way to put the constraints on
the anomalous top-quark interactions not only with a photon and
$Z$-boson, but also with a gluon!

Moreover, such analysis of $ep$-collision (e.g. HERA data) allows
getting the limits on anomalous coupling of the top quark with a
gluon and can be considered as an independent estimation of
$\kappa_g$.

\section*{\bf Acknowledgments}

We are grateful to M.~Mohammadi Najafabadi and O.V.~Zenin for
useful discussions. This work was supported in part by the Dynasty
Foundation and by the Russian Foundation for Basic Research under
Grant \# 08-02-91002-C.

\begin{thebibliography}{**}

\bibitem{Beneke:2000hk}
Beneke~M. {\it et al.}, {\it ``Top quark physics''},
 in {\it ``Standard model physics (and more)
at the LHC''}, G.~Altarelli and M.~L.~Mangano eds., {\it  Geneva,
Switzerland: CERN (2000)} [arXiv:hep-ph/0003033].


\bibitem {fcnc-SM} Grzadkowski~B., Gunion~J.F., and Krawczyk~P.,
 {\it Phys. Lett.}, {\bf B268}, 106 (1991); \\
Eilam~G., Hewett~J.L, and Soni~A., {\it Phys. Rev.} {\bf D44}, 1473 (1991); \\
 Luke~M. and Savage~M.J., {\it Phys. Lett.} {\bf B307}, 387 (1993); \\
Couture~G., Hamzaoui~C., and K{\o}nig~H., {\it Phys. Rev.} {\bf
D52}, 1713 (1995).

\bibitem {dH} Atwood~D., Reina~L. and Soni~A.,
 {\it Phys. Rev.}, {\bf D55}, 3156 (1997)  [arXiv:hep-ph/9609279].

\bibitem {SUSY} Yang~J.M., Young~B.-L. and Zhang~X.,
 {\it Phys. Rev.}, {\bf D58}, 055001 (1998)  [arXiv:hep-ph/9705341].

\bibitem {exotic} F.~del~Aguila, J.A.~Aguilar-Saavedra and R.~Miquel, {\it Phys.\
Rev.\ Lett. } {\bf 82}, 1628  (1999) [arXiv:hep-ph/9808400].

\bibitem{Abe:1997fz}
Abe~F. {\it et al.}  [CDF Collaboration], { \it Phys.\ Rev.\ Lett.\ }
 {\bf 80}, 2525 (1998).

\bibitem{fcnc-HERA}
Wolf~G., [arXiv:hep-ex/0105055]; \\
Alan~A.T. and Senol~A., {\it Europhys.\ Lett.\ } {\bf 59}, 669
(2002) [arXiv:hep-ph/0202119];\\
Dannheim~H.  [H1 Collaboration], [arXiv:hep-ex/0212004].

\bibitem{Aktas:2003yd}
Aktas~A.  {\it et al.}  [H1 Collaboration], {\it Eur. Phys. J.}
{\bf C33}, 9 (2004) [arXiv:hep-ex/0310032].

\bibitem{Chekanov:2003yt}
Chekanov~S. {\it et al.} [ZEUS Collaboration], {\it Phys. Lett.}
{\bf B559}, 153 (2003)   [arXiv:hep-ex/0302010].

\bibitem{lep-2}
Obraztsov~V.F., Slabospitsky~S.R., and Yushchenko~O.P., 
{\it Phys.\ Lett.\ } {\bf B426}, 393 (1998)
[arXiv:hep-ph/9712394];
\\
Heister~A. {\it et al.}  [ALEPH Collaboration],
{\it Phys.\ Lett.\ } {\bf B543}, 173 (2002) [arXiv:hep-ex/0206070]; \\
Abreu~P. {\it et al.}  [DELPHI Collaboration],
{\it Phys.\ Lett.\ } {\bf B446}, 62 (1999) [arXiv:hep-ex/9903072]; \\
Abdallah~J. {\it et al.} [DELPHI Collaboration],
{\it  Phys.\ Lett.\ } {\bf B590}, 21 (2004)  [arXiv:hep-ex/0404014];
\\
Achard~P. {\it et al.}  [L3 Collaboration], {\it Phys.\ Lett.\ }
{\bf B549}, 290 (2002) [arXiv:hep-ex/0210041]; \\
Abbiendi~G. {\it et al.}  [OPAL Collaboration], {\it Phys.\ Lett.\ }
{\bf B521}, 181 (2001) [arXiv:hep-ex/0110009].

\bibitem{Tevatron}
Abazov~V.M. {\it et al.} [D0 Collaboration] {\it Phys.\ Lett.\ }
{\bf 99}, 191802 (2007) [arXiv:hep-ex/0702005].

\bibitem{SMtopprod}
Stelzer~T., Sullivan~Z. and Willenbrock~S., {\it Phys. Rev.} {\bf
D 56}, 5919 (1997) [arXiv:hep-ph/9705398]; \\
Moretti~S. and Odagiri~K., {\it Phys. Rev.} {\bf D 57}, 3040
(1998) [arXiv:hep-ph/9709435].


\bibitem{tait}
 Malkawi~E. and Tait~T.,
  {\it Phys.\ Rev.\ } {\bf D54} (1996) 5758   [arXiv:hep-ph/9511337]
\\
  Tait~T. and Yuan~C.~P.,
 {\it  Phys.\ Rev.\ } {\bf D55} (1997) 7300   [arXiv:hep-ph/9611244].


\bibitem{toprex}
Slabospitsky~S.R. and Sonnenschein~L., {\sf ``TopReX generator
(version 3.25): Short manual''}, {\it Comput.\ Phys.\ Commun.\ }
{\bf 148}, 87 (2002) [arXiv:hep-ph/0201292].

\bibitem{belyaev}
Belyaev~A. and Kidonakis~N., {\it Phys. Rev.} {\bf D65}, 037501
(2002).


\end{thebibliography}
\end{document}